\documentclass[12pt]{iopart}
\usepackage{amsbsy}
\usepackage{graphicx, epsfig, amssymb, mathdots} %include figure files
\usepackage{bm} %include bold math: \bm{} creates bold letters in math mode

\usepackage[linktocpage]{hyperref}
\usepackage[caption=false]{subfig}
\usepackage[usenames]{color}%\usepackage{ulem}
\usepackage{url}
\usepackage{color}
\newcommand{\be}{\begin{equation}}
\newcommand{\ee}{\end{equation}}
\newcommand{\beq}{\begin{eqnarray}}
\newcommand{\eeq}{\end{eqnarray}}

\begin{document}

\title{Love in Extrema Ratio}
%\\{\color{red} Circulation restricted}}

%\input{AuthorList.tex}

\author{Paolo Pani\footnote{paolo.pani@uniroma1.it~(corresponding author)}, Andrea 
Maselli\footnote{andrea.maselli@roma1.infn.it}}
\address{Dipartimento di Fisica, ``Sapienza'' Universit\`a di Roma \& Sezione INFN 
Roma1, Piazzale Aldo Moro 5, 00185, Roma, Italy}

\begin{abstract}
The tidal deformability of a self-gravitating object leaves an imprint on the gravitational-wave signal 
of an inspiral which is paramount to measure the internal structure of the binary components.
We unveil here a surprisingly unnoticed effect: 
in the extreme-mass ratio limit the tidal Love number of the central object (i.e. the quadrupole moment induced by the 
tidal field of its companion) affects the gravitational waveform at the leading order in the mass ratio.
This effect acts as a magnifying glass for the tidal deformability of supermassive objects but was so far neglected, 
probably because the tidal Love numbers of a black hole (the most natural candidate for a compact supermassive 
object) are identically zero. 
We argue that extreme-mass ratio inspirals detectable by the future LISA mission might place constraints on the tidal 
Love numbers of the central object which are roughly $8$ orders of magnitude more stringent than current ones on neutron 
stars, potentially probing all models of black hole mimickers proposed so far.
\end{abstract}

\vfill
% \vspace{3cm}
\begin{center}
\begin{small}
Essay written for the Gravity Research Foundation 2019 Awards for Essays on Gravitation
\end{small}
\\
%  \begin{footnotesize}
%   Submission date: \today
%  \end{footnotesize}
\end{center}

\maketitle

% 
% We will make these Awards on May 15, 2019 for the best and most well-written essays about 
% gravitation, its theory,
% applications or effects. Essays should be 1500 words or fewer excluding abstracts and a 
% small number of equations,
% diagrams, tables and references. The subject matter may or may not be original research. 
% The essay competition is not
% intended to replace a research journal where the detailed results of original research are 
% submitted. Essays should not give
% lengthy detailed mathematical calculations nor detailed descriptions of an experimental 
% setup. Essay ideas should be self-
% contained and understandable - not dependent on reading other documents.

%%%%%%%%%%%%%%%%%%%%%%%%%%%%
% \section{Introduction}
%%%%%%%%%%%%%%%%%%%%%%%%%%%%

\hskip 0.2\textwidth
\parbox{0.8\textwidth}{
\begin{flushright}
{\small 
\noindent {\it ``Love all, trust a few, do wrong to none.''}\\
William Shakespeare, All's Well That Ends Well
}
\end{flushright}
}
% Love is just a word, but you bring it definition. Eminem
% You can’t blame gravity for falling in Love. Albert Einstein

\vspace{1cm}

Gravitational-wave~(GW) measurements of the tidal deformability of neutron stars~\cite{Flanagan:2007ix} 
--~through the so-called tidal Love numbers~(TLNs)~\cite{PoissonWill}~-- provide one of the most accurate 
tools to date to probe the microphysics of the neutron-star interior~\cite{Abbott:2018exr,De:2018uhw}.
It has been recently realized that tidal effects in coalescing binaries can also be used to
distinguish black holes~(BHs) from other ultracompact objects~\cite{Wade:2013hoa,Cardoso:2017cfl,Sennett:2017etc}. 
A remarkable result in classical General Relativity is that --~owing to the one-way nature of the event horizon~-- the 
TLNs of a BH are identically zero~\cite{Binnington:2009bb,Damour:2009vw,Gurlebeck:2015xpa,Pani:2015hfa,Pani:2015nua,Landry:2015zfa}, 
whereas those of an ultracompact horizonless object are small but 
finite~\cite{Pani:2015tga,Porto:2016zng,Cardoso:2017cfl}.
Beside posing an intriguing problem of ``naturalness'' in Einstein's theory~\cite{Porto:2016zng}, this precise 
cancellation provides also an opportunity to test the BH paradigm: measuring a non-vanishing TLN with measurements 
errors small enough to exclude the null case would provide a smoking gun for the 
existence of new species of ultracompact massive objects.

The impact of the tidal deformability on the GW signal from a binary coalescence has been so far studied mostly in the 
case of comparable masses. This choice is well motivated for neutron star binaries, but it might be too 
restrictive in the context of tests of the nature of dark compact objects, especially because future GW observations 
are expected to unveil binaries with mass ratios departing significantly from unity.
With this motivation in mind, here we explore the following question: \emph{how much does the tidal deformability 
affect an extreme-mass ratio inspiral~(EMRI)~\cite{Berry:2019wgg} when the massive 
central object has non-vanishing TLN?}

Let us consider a non-spinning compact binary, with masses $m_i$ ($i=1,2$), total 
mass $m=m_1+m_2$, and mass ratio $q=m_1/m_2\geq1$. At leading post-Newtonian order, the correction to the 
\emph{instantaneous} GW phase due to the tidal deformability of the binary components 
reads~\cite{Flanagan:2007ix,PoissonWill} (we use $G=c=1$ units)
\begin{equation}
\phi_{\rm tidal}(f)=-\frac{117}{8}\frac{(1+q)^2}{q}\frac{\Lambda}{m^{5}} v^{5}\,, \label{phaseTidal}
\end{equation}
where $v=(\pi m f)^{1/3}$ is the orbital velocity, $f$ is the GW frequency,
% ${\cal {\cal M}}=\frac{(m_1 m_2)^{3/5}}{m^{1/5}}$ is the chirp mass,
%%%
\begin{equation}
 \Lambda=\frac{1}{26}\left(\left(1+{12}/{q}\right)\lambda_1+(1+12 q)\lambda_2\right)\,,
\end{equation}
%%%
is the weighted tidal deformability, whereas $\lambda_i=\frac{2}{3} m_i^5 k_i$ and $k_i$ are the tidal 
deformability and the (dimensionless) TLN of the $i-$th object, respectively. These can be defined in terms of the 
quadrupole moment $Q_{ab}^{(i)}$ of the $i$-th object induced by the tidal field $G_{ab}^{(j)}$ produced by its 
companion, namely
%%%
\begin{equation}
 Q_{ab}^{(i)} = \lambda_i G_{ab}^{(j)} \sim \lambda_i \frac{m_j}{r^3}\,,  \qquad i\neq j \label{QTLN}  
\end{equation}
%%%
where $r\sim m v^{-2}$ is the orbital distance. For a typical neutron star, $k_i\approx 1000$ and 
$\Lambda/m_i^5\approx 600$, the exact values depend on the equation of state. As a rule of thumb, the more 
compact an object, the smaller its TLN, so much so that a BH has $k_{\rm BH}=0$.

It is enlightening to expand Eq.~\ref{phaseTidal} in the extreme mass-ratio limit.
Let us first consider the standard case in which the central object is a BH, so that $k_1=0$.
In such case the first nonvanishing contribution is
%%%
\begin{equation}
 \phi_{\rm tidal}(f)\sim -\frac{9}{2} k_2 v^5 \frac{1}{q^3}+...\,, \qquad q\gg1~~~~(k_1=0)
\end{equation}
%%%
which is proportional to the TLN of the \emph{small} companion, and is suppressed by a $q^{-3}$ factor.
% relative to Eq.~\ref{phaseTidal1}.
Therefore, for a typical EMRI with $q\approx 10^6$, the above term is negligibly small.

On the other hand, when $k_1\neq0$, the tidal phase \emph{grows linearly} in $q$,
\begin{equation}
\phi_{\rm tidal}(f)\sim -\frac{3}{8} k_1 v^5 q+...\,, \qquad q\gg1 \label{phaseTidal1}
\end{equation}
and is proportional to the TLN of the central object. 
% Clearly the 
It is remarkable that in this case the tidal phase enters at the leading order in the mass ratio just like the 
ordinary radiation-reaction term,
%%%
%
\begin{equation}
\phi_{N}(f)\sim \frac{3}{128} v^{-5} q+...\,, \qquad q\gg1   \label{phaseN}
\end{equation}
although the latter dominates at large binary separation, owing to the different scaling with the orbital 
velocity.
The tidal phase contribution in Eq.~\ref{phaseTidal1} has the same scaling with $q$ as the spin-induced quadrupolar 
deformations~\cite{Barack:2006pq} and, as long as $k_1\gtrsim 1/q$, it is even \emph{larger} 
that the first-order correction due to the conservative part of the self force (i.e., the self-interaction of 
a test-particle with its own gravitational field~\cite{Barack:2009ux}), the latter being suppressed by a factor 
$1/q$ relative to Eq.~\ref{phaseN}.

The above intriguing result case be explained as follows. The GW phase can be obtained by solving for
%%%
\begin{equation}
 \frac{d^2\phi(f)}{df^2}=\frac{2\pi}{\dot E}\frac{dE}{df}\,, \label{phaseeq}
\end{equation}
%%%
where $E$ is the binding energy of the binary and $\dot E$ is the energy flux emitted in GWs. The TLNs enter both in 
conservative piece, $E(f)$, and in the dissipative piece, $\dot E$. To the leading-order in post-Newtonian 
theory~\cite{Vines:2011ud}
%%%
\begin{eqnarray}
      E(f) &=&  -\frac{m_1}{2(1+q)}v^2\left(1-\epsilon_c \frac{6q(k_1 q^3+k_2)}{(1+q)^5} v^{10}\right)\,, \\
 \dot E(f) &=&  -\frac{32}{5}\frac{q^2}{(1+q)^4}v^{10}\left(1+\epsilon_d \frac{4 \left(q^4 (3+q) k_1+(1+3 q)
   k_2\right)}{(1+q)^5} v^{10} \right)\,, \label{Edot}
\end{eqnarray}
%%%%
where $\epsilon_c$ and $\epsilon_d$ are just book-keeping parameters for the correction to the conservative and 
dissipative term, respectively. 
% Notice that, in the extreme-mass ratio limit, the correction to the energy is 
% suppressed by $1/q$ relative to the Newtonian term, whereas the correction to the flux is constant and proportional to 
% $k_1$.
%
Plugging this into Eq.~\ref{phaseeq} and solving at the leading order in the corrections, one finds
%%%%
\begin{equation}
 \phi(f)=\phi_N(f)\left(1-16\epsilon_d k_1 v^{10}\right)\,, \qquad q\gg1 \label{phaseT}
\end{equation}
%%%%
whereas the correction coming from the conservative term is subleading. It is straightforward to 
check that the above equation yields Eq.~\ref{phaseTidal1}.
Thus, the enhancement of the tidal effect in the waveform is due to the contribution of the TLNs to the energy flux. 
When $k_1\neq0$, this term is not suppressed by any power of $1/q$ relative to the leading-order term, as evident 
from 
Eq.~\ref{Edot}.
In turn, this result can be obtained straightforwardly by using Eq.~\ref{QTLN} and the quadrupole formula, $\dot E\sim  
\partial_t^3 Q_{ij}^{\rm tot} \partial_t^3 Q_{ij}^{\rm tot}$, where $Q_{ij}^{\rm tot}$ is the total quadrupole of the 
binary.

Let us now discuss the phenomenological implications of this enhancement. 
We consider an EMRI up to the innermost stable circular orbit~(ISCO) of the central object.
The total GW phase accumulated between 
$f_{\rm min}$ and $f_{\rm max}\sim f_{\rm ISCO}=\frac{1}{6\pi\sqrt{6}m_1}\gg f_{\rm min}$ due to the tidal 
deformability is simply
%%%
\begin{equation}
 \phi_{\rm tidal}^{\rm tot}=-\frac{k_1}{96\sqrt{6}} q\approx -0.004 k_1 q \,. \label{phaseGW}
\end{equation}
%%%
If for instance $q=10^7$, by requiring a detectability threshold $\phi_{\rm tidal}^{\rm tot}>1\,{\rm rad}$, we find that 
the effect might be measurable even for TLNs as small as $k_1\approx2\times 10^{-5}$. 

This bound is quite impressive at least for two reasons. First of all, it suggests that the future LISA 
mission~\cite{Audley:2017drz}, which is expected to detect few to thousands EMRIs per 
year~\cite{Babak:2017tow,Berry:2019wgg}, could 
set constraints on the TLN of the central object which are approximately \emph{$8$~orders of magnitude smaller} 
than LIGO's current measurements on the tidal deformability of a neutron star~\cite{Abbott:2018exr,De:2018uhw}. 
Furthermore, in the most extreme models of horizonless compact objects~\cite{Cardoso:2017cfl} --~some of which 
predicting quantum corrections at the horizon scale~-- the TLNs are of the order $k_1\approx 10^{-3}$, well above the 
bound estimated in Eq.~\ref{phaseGW}. Finally, it is easy to show that the relative measurement errors on $k_1$ 
scale as $q^{-1/2}$ in the high signal-to-noise ratio limit. Thus, EMRIs detectable by LISA might provide the 
ultimate tests for exotic compact objects (see Ref.~\cite{Cardoso:2019rvt} for a review).

Besides the aforementioned bounds on BH mimickers, the enhancement of the tidal phase in an EMRI might be 
relevant if primordial BHs with masses $m_2\approx 10^{-4}\,M_\odot$ exist in nature. These objects might form an EMRI 
around a neutron star and pass through LIGO's band in less than a year before plunge. In such case the tidal 
phase would provide an unparalleled way to constrain the neutron-star equation of state through a 
measurement of the TLN at the level given by Eq.~\ref{phaseGW}. Unfortunately, the event rates for 
neutron-star capture of primordial BHs seem extremely small in this mass range~\cite{Capela:2013yf}.

%Closing
Our derivation is based on a low-velocity expansion of the field equations, and the 
post-Newtonian series converges poorly in the large-mass ratio limit, at least in its 
dissipative sector~\cite{Fujita:2011zk}. Therefore, Eq.~\ref{phaseT} cannot be 
used for a rigorous parameter estimation, but it should nonetheless provide the correct order of magnitude of 
the effect of the tidal deformability of the central object.
We conclude that it would be very important to incorporate the tidal deformability terms consistently in an 
extreme-mass ratio expansion of Einstein's field equation for binary systems beyond post-Newtonian theory.

All in all, Love might be at play even in extreme encounters.

% \pp{Time to ISCO starting at $f_{\min}\ll f_{\rm ISCO}$, ti torna?:}
% %%
% \begin{equation}
%  T_{\rm ISCO}\sim 0.2 \frac{q}{10^4} \left(\frac{1.4}{m_1}\right)^{5/3} \left(\frac{10\,{\rm Hz}}{f_{\rm 
% min}}\right)^{8/3}\,{\rm yr}
% \end{equation}

% \pp{TO BE REMOVED:}
% 
% More rigorously, a simple 
% Fisher-matrix analysis --~valid only for large signal-to-noise 
% ratios~(SNRs)~-- yields the following relative error \am{domani mattina passo da te}:
% %%%
% \begin{equation}
%  \frac{\Delta k_1}{k_1}\sim 1.81\times 10^3 \left(\frac{d}{1\textnormal{Gpc}}\right)\left(\frac{10^{-3}}{k_1}\right) 
%  \left(\frac{10^5}{q}\right)^{1/2}\left(\frac{10^6 m_\odot}{m_1}\right)^{5/2}
%  \label{constraints}
% \end{equation}
% %%%
% where we assumed the typical sensitivity curve of the future LISA mission~\cite{Audley:2017drz}, 
% which is expected to 
% detect few to thousands EMRIs per year~\cite{Babak:2017tow}.

%%%%%%%%%%%%%%%%%%%%%%%%%%%%%%%%%%%%%%%%%%
\section*{Acknowledgments}
%%%%%%%%%%%%%%%%%%%%%%%%%%%%%%%%%%%%%%%%%%
We are grateful to Tiziano Abdelsalhin, Leor Barack, Enrico Barausse, and Scott Hughes for interesting comments on the 
draft.
PP acknowledges financial support provided under the European Union's H2020 ERC, Starting 
Grant agreement no.~DarkGRA--757480.
The authors would like to acknowledge networking support by the COST Action CA16104 
and support from the Amaldi Research Center funded by the MIUR program "Dipartimento di 
Eccellenza" (CUP: B81I18001170001).

%
%%%%%%%%%%%%%%%%%%%%%%%%%%%%%%%%%%%%%%%%%%
\section*{Bibliography}
%%%%%%%%%%%%%%%%%%%%%%%%%%%%%%%%%%%%%%%%%%
%
%%%%%%%%%%%%%%%%%%%%%%%%%%%%%%%%%%%%%%
\bibliographystyle{myutphys}
\bibliography{refs}

\end{document}